\documentclass{article}

\newcommand{\set}[1]{\{#1\}}

\title{Journalistic Voting System's Effects %
on Election Security Threats and Gerrymandering}
\author{Lucius Schoenbaum}
\date{\today}

\begin{document}

\maketitle

\begin{abstract}
The Journalistic Voting System is a proxy voting system %
in which journalists are delegated the task of voting %
on behalf of individual voters in a western-style democracy. %
We introduce the Journalistic Voting System %
and discuss its potential advantages and potential problems. %
In particular, we discuss its advantages to individuals %
in the system (voters, journalists, and politicians) %
and we discuss its effects relative to %
several widely discussed threats to election security, namely: %
cybersecurity, social media, big data, artificial intelligence (AI), and gerrymandering. %
The Journalistic Voting System is modeled on a predecessor system, %
called the Valence Voting System, which is reviewed. %
\end{abstract}

\section{Introduction}\label{s.i}

The Valence Voting System (first proposed in \cite{VV18}) %
and the Journalistic Voting System (introduced below) are voting systems %
developed in response to present threats to election security in western-style democracy, %
and the closely related problem of gerrymandering. %
In particular, consider the following list of interconnected problems, or ``threats'': %
\begin{enumerate}
\item 
computer-generated and/or doctored images, audio, video, and text (``Artificial Intelligence'') 
\item 
the unprecedented amount of data currently available to reveal human behavior, activity, vulnerabilities, and preferences (``Big Data'') 
\item 
effortless, instant, essentially frictionless communication via large private, profit-based social media platforms who help decide %
what voters see and do not see in the news media and via politicized social media artifacts, 
sometimes with the aid of privately established quasi-legal frameworks (``Social Media'') 
\item 
progressively lowering impedance in cost and difficulty to the construction of highly scalable systems %
(``Bots'', advertising campaigns, blogging networks, etc.) 
\item 
cross-border communication networks and platforms spanning geopolitically competing nation-state actors (election ``Hacking'', election ``Meddling'') 
\item 
widely available capacity to construct computer-aided, goal-optimized political boundaries and maps (``Gerrymandering'') 
\end{enumerate}
Each of these are well-explored topics, and the interconnected threats they pose is widely acknowledged. %
If the reader is not involved in them directly in some capacity, %
she is likely familiar with them and has likely overheard stories about them second-hand and through the media. %
Therefore, although these are important topics that motivate what follows, %
we will not review them here. %
In summary, there are certain election security threats which are traceable to %
the practice of western-style democracy in an age of advanced technology %
and abundant, low-cost computational resources. %
These threats are related to one or more of: %
{\bf artificial intelligence, big data, social media, and cybersecurity.} %
In addition, {\bf gerrymandering} presents an ongoing %
concern for fair representation in the practice of western-style democracy. %

\section{Valence Voting System}\label{s.vv}

The Valence Voting System is a system which aims to solve the problem of gerrymandering %
by allowing voters to choose their districts themselves. %
Now we describe the Valence Voting System. %

Let $\set{V}$ be the set of voters. %
Each voter $V$ is assigned (via address/location) a {\em cell} $C$ out of the total set of cells $\set{C}$. %
We can think of a cell $C$ as being an area of the total geography that contains 10,000 people, on average. %
The cells partition the total geography, and they do not overlap. They are determined %
by census officials and they are assumed to be simple and nondescript. For example, they might be %
squares divided by lines of latitude and longitude, %
except where they encounter natural boundaries, such as bodies of water. %
Each voter $V$ has some {\em valence}, a certain quantity that each voter %
may distribute to politicians during an election. %
It is conventional to multiply the valence by 100, so that %
each voter $V$ has a valence value of 100. %
This implies that, in our rough picture, %
a cell has, on average, 1,000,000 total valence. %

Let $\set{P}$ be politicians (representatives and candidates). %
The Valence Voting System can only apply to representatives, not to unitary office-holders (such as governors and mayors). 
In the course of an election cycle, each politician $P$ can stand for re-election, %
or stand and contest a seat. %
However, unlike in more familiar systems, a politician does not choose a district to stand in: %
every politician stands simultaneously in every ``district'', that is, in every cell, %
simultaneously. %
Of course, in theory, the politician will select a region or area %
in which to focus his or her campaign. %

There is an election day, as usual, %
when the voting takes place. %
On election day, the voter $V$ ``votes'' in the following manner: %
he or she, in exercise of their own free choice, %
distributes their valence of 100 to as many politicians as he or she chooses, %
out of the total number $\set{P}$ who have stood. %
Via technology, this entire list can be distributed to the entire population. %
A ballot of that size is possible %
by technological enhancement of the traditional paper ballot, %
or by a paper ballot with a large write-in column. %

The winners of the election in each cell are determined by a valence {\em cutoff} %
determined in each cell. %
For simplicity, assume that the valence cutoff is simply set at 1\%. %
Thus, if each cell contains 10,000 people, %
then a politician is awarded valence by the cell if that valence is larger than %
1\% of 1,000,000, or 10,000 valence. That is, 1\% or more of each person's valence in the cell, on average. %
If valence is described as currency, say \$100, then %
one dollar's worth of valence from each member of the cell awarded to politician $P$ %
is enough for $P$ to acquire representation during the next term. %
It is theoretically possible to award negative valence, which would cancel out positive valence. %
This would inhibit ``controversial'' representatives, but for discussion's sake, assume that valences are always positive %
since otherwise the voting tally is more complicated. %

A second way in which a politician can ``win'' is via {\em special} representation. %
This means that the politician %
wins over (say) 1\% of the total valence available in {\em all} cells. %
In that case, the politician is elected as a {\em special} representative. %
This exception is allowed to give representation to groups that are distributed geographically, %
but that due to sparsity do not form a significant fraction in any specific cell. %
In this case, cells are notified and an exception is made to that candidate, %
if he or she is below the cutoff. %

Valence is {\em inherited} by politicians, in the sense that %
a politician does not just acquire a ``seat'', but rather, acquires the valence %
awarded by his or her voters. %
This valence is preserved throughout his or her term as measure of the weight %
of his or her vote in the representative body. %
Thus, if a politician wins a total valence of, say, 300,000,000, %
then that politician will have twice as much voting power %
as another politician who wins a total valence of 150,000,000. %
If we divide the politician's valence by the total population in all of the cells, %
then politicians, too, would have a valence (roughly) out of 100. %

Valence that is distributed to candidates that do not win %
must be distributed in a follow-up procedure, %
which guarantees that each cell is fairly represented %
in the ultimate body's voting.  %
One way in which this can be done is the following. %
A list of the politicians who avoided the cutoff, and the special (exceptional) representatives is prepared, %
and voters who still have ``open'' valence are contacted, and asked to distribute their valence %
among the candidates on that list. %
The voter can opt, at this time, to withhold some or all of their remaining valence, if he or she so chooses. %

This describes the main principles of the Valence Voting System. %
The key features of the system are: %
\begin{enumerate}
\item 
Voters are invested with valence by the system, %
and they pass this valence on to a chosen set of representatives, %
who then wield that valence on their behalf in steering the government, drafting and passing laws, etc. %
\item 
Voters choose their representatives and, via the voting process itself, %
districts are formed in an immediate and unanticipated manner. %
A typical district would perhaps have a central peak or plateau, %
and fall off gradually around the edges, %
and show spatial variations in valence corresponding to geographic patterns in the underlying population. %
\end{enumerate}

\subsection{Potential Advantages of the Valence Voting System}\label{s.vv.ad}

In this section, the potential advantages of the Valence Voting System are discussed. %
The discussion is made on a purely theoretical basis. %
Nevertheless, we will speak optimistically in terms of potential, not actual, %
empirically observed advantages. %

Gerrymandering is possible through the techniques of ``cracking'' and ``packing'' %
subgroups within a larger population. %
The technique of ``cracking'' a population is not possible in the present case, %
because an artificially selected line would have an extremely low probability of occuring. %
Even if it did occur, due to the ability of districts to overlap, a minority (even a small minority) %
would have its representation, with a correspondingly lower valence, from each of the ``pieces'' %
of the cracked geographic region, and these would add up to the same total as before. %
Likewise, the technique of ``packing'' would be impossible, %
because the valence of the corresponding representative(s) in ``packed'' cells %
would have an commensurate increase in valence. %
Regardless of how a population is distributed (densely, or sparsely), its representation would occur %
in direct proportion to that population's participation in the process on election day and afterwards. %

In addition, because a voter's valence can be divided among different representatives, %
voters would be able to represent their own identity more faithfully in the representative body. %
For example, a voter who is, in all respects except one, exactly like the other members %
of a cell, can separate his or her valence, reserving a portion of it in order to express %
that which is singular and unique in his or her identity, %
possibly in coordination with a group in the wider population, %
whose aim is to achieve representation in the representative body. %
In other words, it is not the largest common denominator, %
but instead the {\em intersectional identity} of a person, %
that is, the different facets of the person %
that together form that person's particular identity, %
that is politically expressed. %
A lamentable choice between ``the lesser of two evils'', %
a commonly-heard criticism in a two-way, first-past-the-post contest, could be averted. %

In addition, politicians would be incentivized to %
act in ways that are more faithfully representative of the people's will. %
There would be no concept, in the Valence Voting System, %
of ``barely winning'' an election. 
An election that is barely ``won'' %
would be reflected in that politician's reduced valence, relative to another politician %
who ``won'' their seat by a wide ``margin'', to use the phrase from modern political campaigning. %
This would disincentivize the worst types of abuse of political office, %
and some of the familiar antics of political campaigning: %
the most effective campaign of all would be to simply be popular and win high approval. %

Nor is there any notion of ``winner takes all'' and ``loser goes home'' %
in a Valence Voting System. %
Candidates do not compete directly over a scarce resource, but instead, %
they compete in a way that tends to disincentivize direct attacks, and encourage %
greater cooperation and tolerance. %
A candidate who attacks another candidate would face the possibility that %
some of the other candidate's supporters are the very same voters %
who are also supporting him. %
When a campaign is over, %
popular representatives would all ``win'', %
and it would be likely that in most cases one's time of exit from one's role as a representative would %
come at a time of one's own choosing. %

\subsection{Problems with the Valence Voting System}\label{s.vv.prob}

Recall that the discussion %
takes place prior to any kind of empirical testing. %
With that in mind, %
there are some basic problems with the Valence Voting System. %
\begin{enumerate}
\item
The lack of jurisdictions leads to a managerial and operations bottleneck %
that might lead to a unwieldy, dysfunctional government, with unpredictable dynamics. %
\item 
It will be more easy for purists to operate politically, and these purists might %
be able to create gridlock and sow further division. %
\item 
The size of the representative body is not fixed, %
and it could be expected to be large enough to make the functioning of the body impractical. %
\end{enumerate}
We can easily make a rough estimate of the size, %
if we notice that using the example figures above, %
the largest number of representatives a cell can have occurs %
when each member of a cell votes for exactly the same 100 representatives. %
So an upper bound on the size of the representative body, given the example figures used here, would be %
$N \times 100 \div 10,000 = N/100$, where $N$ is the size of the population. %
For the population of the USA, this figure would be 3.3 million. %
On the other hand, the size of the representative body could, in theory, %
dwindle to the most minimum positive size of 1, if all voters awarded all of their valence to the same representative. %

Broadly speaking, the basic problem with the Valence Voting System is %
that the dynamics of the elected body are too permissive and too weakly controlled. %
The act of governance is challenging and requires establishing %
reliable priors on which to base a course of action, %
and a managable number of principal individuals %
involved in a decision or action. %
Governing individuals have to build a %
working relationship with other governing individuals, in order to %
work through differences and come to an agreeable path forward to avoid conflict %
and the real threat of a managerial disaster or snafu. %
The slow and complex work of passing written documents representing the collective will, %
and bills that go on to become laws, %
requires the ability to predict the course and outcome of procedural maneuvers. %
This is made far more challenging, if not impossible, %
if an extremely high number of individuals (on the order of $10^4$, $10^5$, or higher) 
is able to directly influence the outcomes of votes. %

A second problem with the Valence Voting System %
is that it is not empirically tested. %
We simply do not know how it would perform at any scale, %
and our considerations are only approximate at best. %
However, we draw the conclusion that the Valence Voting System %
has both compelling advantages, and major disadvantages. %

\section{Journalistic Voting System}\label{s.jd}

The Journalistic Voting System %
is a response to the problems with the Valence Voting System. %
It is a system in which a proxy voting layer is enacted between politicians and voters. %
We now describe the Journalistic Voting System. %

In the system, there is a set of voters~$\set{V}$ who 
live in at least one, probably more than one, districts from a set~$\set{D}$. %
Districts are the legally defined geographic subdivisions of the total geography. %
Under this understanding, municipal boundaries, county lines, state lines, and so on %
(any jurisdiction where an election might take place) %
would each define a district~$D$. %
Districts may overlap, and indeed, they are expected to overlap and be nested. %
At the lowest level, there might be counties, and above this, there might be cities, %
states, and additional higher levels, as the case might be, up to a maximum all-encompassing district size. %

Next, there are politicians $\set{P}$, including candidates, representatives, %
and officials. An {\em official} is a unitary office-holder, such as a governor or mayor. %
Finally, there are {\em journalists}, the set of which is~$\set{J}$. %
A voter $V$ and a politician $P$ are intended to correspond to the familiar, %
time-honored concepts. %
What we mean by a journalist $J$ probably requires some explaining. %

What first of all defines a journalist $J$ is that he or she has registered as a journalist. %
Much like for a politician running for office, %
this registration requires fulfilling some preliminary criteria %
(for example, a minimum number of signers of a petition). %
An initially registered journalist is awarded a small stipend, but once a journalist has accumulated %
enough valence (to be defined shortly), he or she is allowed to collect an annual salary %
provided by the state, %
essentially becoming a journalist working full time. %
Thus, the journalist $J$ is essentially a public employee, or a {\em public journalist}. %
Once a journalist is registered, he or she is subject to regulations similar to anyone %
who holds a public office, for instance, rules concerning meetings with foreign agents, %
rules concerning solicitation or acceptance of bribes, and so on. %
Like any employee or public servant, a journalist can quit at any time to pursue %
other goals. %

The reason that a journalist is referred to as a {\em journalist} %
is that he or she would be expected, and to some extent, required to function more or less %
precisely as a private journalist does. %
(By {\em private journalist,} we mean the usual notion of journalist.) %
This means that he or she would live as a public figure, %
make regular appearances in the media, %
follow politics as well as other journalists (for tips and updates), %
and post regular updates on social media platforms. %
These avenues are the primary ways in which the journalists %
builds and maintains his or her connection with those %
who rely on him or her for news, updates, and analysis. %
This, in turn, is converted into the public journalist's political influence, %
that is, his or her valence, and it also propels his or her career as a journalist. %

An important concept, borrowed from the Valence Voting System, is %
the journalist's valence, or {\em influence}. %
This is the total amount of valence that the journalist has collected at any given time. %
Much like a follower count, or subscriber count, %
it is manifestly a number (required to be made public) %
indicating the journalist's valence: the total $V(J)$, and by district, $V(J | D)$. %
The valence is awarded by subscribers and followers %
via a publically maintained app or web portal, %
perhaps similar to the IRS tax portal in the United States. %

As in the Valence Voting System, %
a voter (once they are of voting age) is awarded a quantity of valence, %
which is 100 in total, by convention. %
They control this valence throughout their lifetime. %
This control is based on the freedom to make real-time updates %
similar to actions taken in an online financial account. %
At any time, voters can modify their valence distribution by visiting the portal. %
Valence can be withheld (remain ``open'') or distributed to registered journalists, %
and only to them, in any fashion chosen by the voter. %
Borrowing another idea from the Valence Voting System, %
{\em there is no geographic jurisdiction to a journalist}. %
Any voter can award valence to any journalist. %
There is no cutoff, as in the Valence Voting System. %
Once the valence is awarded, it sticks, %
until either the journalist or the voter is removed from the system, %
or the voter electively removes or modifies it (which he or she can do at any time). %

The Journalistic Voting System also differs from the Valence Voting System in the following way. %
In the simpler Valence Voting System, %
valence is the mechanism that determines {\em how much voting power} a politician has. %
In the Journalistic Voting System, valence is the mechanism %
by which it is determined both %
how much voting power a {\em journalist} has, and {\em which elections a journalist votes in}. %
It is not passed on to the politician. %
Once the system reaches the stage of an election for political office, %
the system is exactly like a familiar democratic process, with the prominent exception that %
{\em it is a pool of journalists, not the voters, who vote in the election}. %
They vote with a weighting that directly depends on their valence {\em within the district where the election is contested}. %
As we will see below, this %
is done in order to ensure that accountability of journalists persists %
all the way up to the day of the election. %
This pool is directly determined by the valences of these journalists, %
again {\em within the district where the election is contested.} %
This can be implemented in the manner that we now describe. %

For a journalist, there are two values of merit relative to a district: %
first, the {\em lower threshold} %
$$
V_L = \frac{V(J | D) }{V(D)}
$$
that is, the journalist $J$'s valence, relative to the total amount of valence in the district. %
Then, the {\em upper threshold} %
$$
V_H = \frac{V(J | D) }{ V(J) }
$$
that is, the journalist $J$'s valence, relative to the journalist's total amount of valence. %

Let us see how an election cycle might proceed in the Journalistic Voting System. %
Suppose that an election is approaching in 6-8 months. %
Politicians have fixed their intent to run, and all deadlines to run have passed. %
Now, a grace period begins, which is a period of 1-3 months or so. %
At the time the grace period ends, the valences of all journalists are measured and stored, %
and based on these measurements, %
the threshold levels are set by officials in order to ensure that the %
total number of journalists in the voting pool is sufficient. %
An appropriate setting would be determined by experience, %
but perhaps $V_L > 0.05$, $V_H > 0.25$ (5\% and 25\%, respectively) would be adequate. %
These two levels are set by election officials in order to target a pool of journalists %
that is appropriate given the size of the underlying population of the district, %
with perhaps a minimum of 350~journalists: %
a number sufficiently high that a milieu is created. %
It would be assumed that, with experience, %
this protocol could be firmly set, in order to ensure accountability and fairness. %

If there are not enough journalists falling in the initial threshold range, %
then the first recourse would be to raise the upper threshold, meaning that, effectively, %
higher journalists (journalists with a high valence, relative to the district size) would be ``called down'' to participate. %
This is possible because it is only the journalist's {\em district valence} $V(J | D)$ that would count towards the higher journalist's valence %
in the election. %
If a limit is reached on the higher side (i.e., if there are no remaining higher journalists), %
then the lower threshold is lowered further, meaning that effectively more local %
journalists would be ``called up'' in order to vote ``by committee'', by combining their smaller valences %
into an equivalent larger valence. %
If there is still not enough journalists, then a call for journalists is made, and the election may be postponed. %
However, one optimistically imagines a ``thriving'' system in which incentives %
to be a journalist are such that a spot that opens is quickly filled, %
just as restaurants open on a busy street corner. %
Spots might open regularly as people go on to other things in their lives and careers, %
similar to the path taken by politicians on their way into and out of public service. %

The journalist pool for an upcoming election is thus created %
based on valences that are present at the instant of the closing of the grace period. %
A requirement that arises is that, during the grace period, %
the intent of the voter must be verified. %
This is the time during which the voter who is still undecided %
must finalize their awards of valences. %
By the end of the grace period all voters wishing to count must have confirmed %
in some way to the officials managing the election %
that their valences do in fact reflect the voter's current, up-to-date intent. %
In the implementation being considered, this action, which might %
take seconds and be done online, is the only action technically required from the voter. %

During the peak of election season itself (the final 2-3 months leading up to the day of the election), %
the voters maintain full control over their valence. %
If the valence that a journalist has on election day is different than that %
taken at the end of the grace period, then it is that amount, not the previous amount, %
that is applied in the vote on election day. %
However, it is not necessary for a voter to exercise this freedom; %
a voter who follows his or her ``favorite'' journalists day by day, or perhaps once or twice a week, %
and has faith in their decision-making abilities, can simply passively monitor the developments taking place in the election. %

The period after the end of the grace period %
is thus essentially a matter between the journalists, and the politicians %
who have stood for office. %
Interactions between politicians and journalists %
would amount to the campaign, and they would determine %
who is elected to the office. %
These interactions would most likely involve extended in-person encounters, %
such as wide-ranging public interviews, panels, and debates. %
Certain rules would apply, for instance, private meetings and communications between %
candidates and journalists would be forbidden. %
During this period, voters can go about their business, %
or they can watch the campaign, should it interest them. %
The actual voting on election day would be likely to be anticlimactic in many cases, %
though there might be close votes. At that point, the election cycle would conclude, %
and the system would then await the start of the next election cycle. %

When a journalist votes, his or her vote is public, similar to that of a %
politician who takes a vote in an elected body. In a similar way, %
it would likely be common for the journalist to announce his or her voting intent, %
but it would not be a requirement to do so. After an election, %
if there is popular dissatisfaction %
with a journalist's vote, he or she will be subject to the %
usual type of punishment experienced in social media: %
he or she will experience a drop in follows, subscribers, and valence. %

\subsection{Potential Advantages of the Journalistic Voting System, Regarding Threats}\label{s.jd.threats}

In many respects, the voting system would seem to be primed to deliver very similar outcomes %
to that of a familiar democratic voting system, %
because journalists would be incentivized to closely follow the will of their followers. %
In a previous era, then, this may not have been so significant a difference %
as to be worth the additional complexity of a proxy voter layer. %
However, in light of the election security threats noted in section~\ref{s.i}, %
we can see that there do exist potential advantages to the Journalistic Voting System in the present day. %
As before with the Valence Voting System, we note that the system has not been tested empirically, %
and now we speak optimistically about the potential advantages, not actual or realized ones. %

We can briefly summarize the potential advantages regarding election security threats as follows. %
The additional layer of proxy voters constituted by public journalists %
has the advantage of {\em removing the digital intermediary between politicians and their overseers}, %
and providing {\em greater robustness} to the voting system as a whole. %
Let us go through the threats listed in section~\ref{s.i} one by one %
and see how the Journalistic Voting System could mitigate the threat. %

\subsubsection{Artificial Intelligence}

A journalist proxy layer would be primed, as individuals, on the threats posed by artificial intelligence. %
Since they are journalists, it would not be difficult for election officials to quickly and successfully communicate to them %
that a certain video or artifact is phony. %
They would have the capacity, if skeptical, to ``verify the verifiers'', %
in that they would be able to contact security officials and converse with them, and have all (or at least most) of their %
questions answered, simply by prefacing their contact with, ``I am a public journalist, I just have a few questions...'' %

In the traditional system, voters are unfortunately too numerous to receive this level of attention without placing %
an impossibly heavy burden on the system. %
Journalists, who are numerous but far less numerous than voters, can be granted this privilege. %
Moreover, since they have the crucial bond of trust with voters, they would be able to %
go back to voters and explain to them what has happened. %
It would therefore be much more difficult for a fake video, or other kind of AI-synthesized ``October surprise'' %
to throw a close election in one way or the other, before officials could contain the fallout from the falsified media. %
Assuming that the system worked as intended, and %
voters preserved their sense of trust that the system is designed to effectively protect them, %
it would also be more difficult for an attack, even a sustained attack, to leave lasting damage, %
manifested as voter disaffection. %

\subsubsection{Big Data and Social Media}

It is a widely known and controversial fact that large amounts of data are collected about users %
of modern computer systems, %
particularly by social media platforms, who often package and sell this data as ``advertising analytics'' %
or in similar products. %
This poses a threat in the form of psychological manipulation of users: %
a darker word cloud, for example, streaming in a news feed in the days leading up to an election, %
might lead users to vote against an incumbent, %
and this effect might be all but impossible for voters to detect. %
This poses the threat of allowing a small minority of individuals %
to unilaterally determine outcomes of close democratic elections in a clandestine and systematic way, %
a situation that is fundamentally and recognizably undemocratic. %
A similar scenario, from a range of potential vulnerabilities, %
would be a ploy to trick some voters into staying at home on election day %
by controlling the frequency or accuracy of posts reminding voters to vote. %

Journalists in the Journalistic Voting System can be expected to be highly skilled users of social media. %
They are delegated time, space, authority, and resources to develop a highly accurate, detailed, and contextualized picture of the %
candidates, as well as the stakes of the election. %
They are tasked with voting on behalf of their followers and subscribers as one of the primary duties of their full time job. %
In pursuit of this goal, they would engage politicians and other principal actors %
in the political system. Thanks to the higher cutoff $V_H$ that is implemented, %
these engagements would more likely occur in the four-dimensional world, %
not as streams of images and reports of distant places on their phones and computers. %
Therefore, it is a vastly more challenging task for a bad actor who is %
managing a social media platform %
to influence the voting decision of a journalist, 
than it is for him or her to influence statistical outcomes in a larger voter population. %

This would not, of course, prevent a bad actor from attempting to influence the journalists's %
followers and subscribers in some way. %
That is, the threat is not eliminated entirely. %
However, we can observe that the journalist proxy layer provides a buffer %
or regulator to mitigate the threat that this type of ploy presents in a serious threat scenario, %
because the journalist is able to communicate and explain himself or herself to followers, 
and inspire human loyalty and trust that could counteract and blunt the effects of an attacker that %
has only digital, artificial tools in his or her toolkit. %

\subsubsection{Cybersecurity}

There are at least two separate issues for cybersecurity: %
the security of the election taking place on election day, %
and the security of the portal that voters access to allocate their valences. %
We focus here on the former question, since it is more directly relevant to the voting system. %

If we consider security on election day, %
we can see that arguments can be made that are similar in nature to the arguments we have already made %
about big data, artificial intelligence, and social media. %
Similar to before, we can observe that the system is not made invulnerable, %
but it is made more robust to the threat of a cybersecurity attack. %
In particular, the sheer number of participants on election day is decreased by orders of magnitude. %
Therefore, there are fewer potential points of failure, %
and the burden of cybersecurity would be expected to be far lighter, and cheaper. %
Moreover, the threat of a constitutional crisis caused by a hotly contested election %
outcome {\em on the day of the election} or in its immediate aftermath would be mitigated. %
The vote taking place on election day would be more similar to a vote taken in a representative body %
than to a complicated canvassing operation spanning hundreds, if not thousands, of square miles. %
In many cases, the voting itself would be a necessary formality whose outcome is known well in advance. %
In other cases, the vote might be very close and come down to the final moments. %

Thus, the cybersecurity concern would center around the security of the voter portal. %
This would be a matter best left to experts to consider. %
However, if it is decided that a portal of this type cannot be made secure, %
an in-person system could be opted for instead. %
A breach of such a system would also not suffer from the same level of political sensitivity as that of a direct canvassing system, %
and there would be time to resolve the issue before the potential comes for a political storm or crisis. %
Thus, overall, the system can be expected to have a greater security against a voter system hacking operation. %

\subsubsection{Gerrymandering}

We have put off all discussion of the process of drawing districts in the Journalistic %
Voting System until now. %
It should be clear that a way should exist to address the problem of Gerrymandering, at least partially. %
This is for two reasons. 

First, there is no longer a direct, simple line drawn between political boundaries and the %
probable outcome of an election. %
The chaotic boundaries of the journalistic valence system come between these two, %
and not even a supercomputer who draws such boundaries can possibly %
predict exactly what the ultimate setting of the voter's valences will be, %
and these will play a role in how election outcomes turn out. %
Thus, in the Journalistic Voting System, %
the precise locations of the boundaries drawn in political maps simply matter slightly less. %

Second, there is an obvious choice of arbitrator for these boundaries. %
Politicians, whose careers and fortunes depend in part on the influence of the political map on electoral outcomes, %
must be removed from the responsibility to draw these maps, or they will naturally %
act in their own self interest. %
Journalists, on the other hand, are constrainted only by their valence, that is, their attachment %
to their followers and subscribers. %
Their boundaries would only follow natural ``soft'', constantly changing geographical boundaries in the population. %
Their careers and fortunes are indeed influenced by the final placement of boundaries, %
but their career and personal well being (their survival in the system) is not determined by these boundaries in the same way that those of politicians are, %
and certainly not in any way that they can effectively anticipate. %
However, they are engaged full time with the political landscape %
and are highly sensitive to the issues and the effects that arise around the drawing of the political maps. %
Since, finally, journalists are also accountable to voters, %
it should be they who choose the political map. %

For good measure, here is one possible implementation. %
Periodically, the maps that determine the shapes of districts are updated, %
based on revised population data or other factors. %
The maps for a given jurisdiction are selected via a process like the following: %
For a map determining subdivisions in a given district $D$, %
all journalists with a local valence $V(J | D) > 0$ are awarded a vote weighted by their valence $V(J | D)$. %
Their final votes, the tallies, and the deliberation process are all transparent and open to the public. %
The vote is based on a set of submitted maps and the simple majority map wins. %

\subsection{Potential Advantages of the Journalistic Voting System Regarding Individual Participants}\label{s.jd.ad}

The Journalistic Voting System can be viewed as a response to the problems %
that were observed with the Valence Voting System. %
The basic observation is that the chaotic dynamics %
of the Valence Voting System could be tolerated if they were directed %
not to the final governing body, but instead to a buffer layer of proxy voters %
who themselves elect politicians participating in a more rigid and stable system. %
The second observation is that both the dynamics, as well as the role %
of such an imagined proxy layer, bears a certain resemblance to a group of operators %
in the currently functioning structure of representative democracy, %
namely, the role of the democracy's journalists who have in fact acted as the intermediaries between %
voters and politicians on an informal (or quasi-formal) basis for decades, if not centuries. %

In the familiar democratic system based on elections and canvassing, %
there are only two layers, voters and politicians. %
Simply put, there are two bosses, and each boss is accountable to his or her own boss. %
Journalists, in the familiar democratic system, %
do in fact play a role that is simple to observe. %
They explain the activities of politicians to voters, and sometimes they justify them. %
In other instances, they expose that a politician is doing something that voters should be displeased about, %
effectively holding them to accountability. %
In other words, they do in fact act as {\em buffers} between voters and politicians, %
and they do in fact act as {\em monitors} to politicians. %
In a democratic country, journalistic organizations are usually well regulated, %
in tacit recognition of the fact that journalists do in fact play this functional, quasi-official role. %

The Journalistic Voting System can be considered as replacing this arrangement with %
a similar, but more formal system of checks and balances, as follows. %
First, although it is a somewhat uncomfortable truth, we grant that %
the politicians are, as usual, those entrusted to monitor the behavior of voters, %
usually indirectly via the criminal justice system, the courts, regulations and laws, %
but sometimes directly via emergency declarations and similar powers. %
Next, the voters are entrusted to monitor the behavior of journalists, %
via the public social media platform and the awarding of valence (follows, likes, subscriptions, etc.). %
Finally, the journalists are entrusted to monitor the behavior of politicians, %
via the system of ordinary electoral voting (the ballot box). %
Thus, the Journalistic Voting System %
can be viewed as a triangle of accountability, %
in which each accountable agent is one step removed from its own monitor. %
In effect, each layer acts as a {\em buffer} separating one layer from its overseeing layer, %
and each layer also acts as a {\em monitor} to the opposite layer. %

Now, let us consider what advantages there might be to each role player, %
should the oversight role of journalists be elevated to a more official one, via proxy voting. %
We consider voters, journalists, and politicians, one at a time. %
Let us once again keep in mind that this is a purely theoretical discussion, %
and the proposed system has not been empirically tested. %
\begin{enumerate}
\item 
Voters, in the traditional system, %
are tasked with the important work, in a democracy, of monitoring a distant place that is %
(in most cases) far removed from their lives. %
The activities of individuals involved in the machinery of government are hard to understand. %
Even experts and students who spend years closely observing them find them deeply challenging %
and often very technical. %
They involve complex laws and procedures that are subject to changes and modifications %
that can involve complex negotiations and bargains. %
Voters are tasked with reflecting on all of this, %
evaluating costs and benefits of different policies in a fast and highly interconnected world, %
and passing judgment on the work and the actions of politicians. %

In the Journalistic Voting System, this individual burden is delegated to %
journalists, who are chosen by the voter through the ordinary routines of checking in to social media %
and social networks to find news and updates. Journalists are chosen by the voter %
from an abundant pool, in order to communicate the most serious ideas and events that affect them %
in digestable ways, and in ways that make sense to them. %
Thanks to the selection process, this happens in the way that best addresses their own concerns, %
and best agrees with their own lives, lifestyles, and preferences. %

{\em In the Journalistic Voting System, the voter's burden is eased.} 

\item 
Journalists, in the traditional system, %
are tasked with the important work, in a democracy, %
of broadcasting their work to hundreds, thousands, or perhaps millions of people. %
They face the challenging task of building up and preserving influence, access, name recognition, %
and trust that is the basis of success and effectiveness in their career. %
They also must navigate the complex legal framework that a journalist must accomodate. %
In addition to these challenges, journalists are also tasked with another, %
very different goal, which is to present their users---the very same users %
that the journalist works so hard to earn---with advertisements, %
so that advertisers will provide the funding that the journalist needs to stay in business. %
In many cases the journalist may also face another set of incentives %
due to editorial supervision of their work within their employer organization. %
This can sometimes lead to disagreement and be a source of job dissatisfaction. %

In the Journalistic Voting System, %
a journalist who so chooses can take his or her work into the public sector, %
and become a proxy voter for his or her followers. %
If successful, he or she can do the work of journalism as a public good, and be %
provided with a platform and a well-defined role %
pursuing the many-faceted work of journalism, %
with accountability only to the voters who lend their support to his or her work. %

{\em In the Journalistic Voting System, the journalist's burden is eased.} 

\item 
Politicians, in the traditional system, find themselves in the unenviable position %
of being accountable in ways that present loopholes to their accountability. %
There are many voters who are soundly aware of what they have experienced directly, but %
who lack awareness of the structural features of the politician's world, %
or who are missing the full context of a politician's actions. %
If a politician is able to convince a majority of voters that he or she %
is acting in their interest, even if he or she is not, %
or if a politician can convince a majority of voters that he or she can %
make changes that would be for the better, even if he or she in fact cannot, %
then he or she can derive an electoral advantage sufficient to remain in office, %
even if this is not in the interest of the plurality of voters or the population as a whole. %

Moreover, politicians also face a numerical challenge: a very small number of politicians represent a vast %
number of voters, leaving the politician with no recourse but to despair of ever cultivating a meaningful relationship %
with all the voters he or she represents. %
This incentivizes a tendency to prioritize interactions that maximize efficiency, %
such as a carefully planned media appearance, %
or which present an electoral advantage of some kind, such as the prospect of a significant political campaign donation. %
This dynamic can lead to strains in the relationship between politicians and voters, and inspire alienation, cynicism, and disaffection %
on all sides. %

Furthermore, a politician in the traditional system finds that it is insufficient %
to merely win the support of voters. He or she must go further and {\em motivate} voters %
who will then ``turn out'' on election day, and will not stay home and leave the election for someone else to decide. %
Very often, this involves %
laying aside the meaningful but complex achievements he or she has made, %
and instead stoking the flames of fear and anger over an exaggerated threat from the ``other side'' %
or some imagined alternative if the individual voter does not exercise his or her right to vote. %
In effect, a politician is incentivized to stoke emotions %
in the population that are based on an exaggerated version of reality. %
In some cases, a politician may go so far as to state falsehoods in the course of pursuing this objective, %
if reality is simply not colorful or bold enough to meet his or her immediate needs. %

In the Journalistic Voting System, %
the number of individuals to whom the politician is directly accountable, i.e., journalists, is much smaller %
than the total number of voters that he or she serves. %
The number is likely on the scale of hundreds, or perhaps thousands in some cases. %
This is a small enough number of individuals that the politician can develop a working relationship %
with a large number of them, perhaps even all of them in some cases. %
He or she is able to speak to each individual journalist in person, as well as to groups of them %
in intimate public gatherings, having the chance to explain himself or herself %
in detail and at length. %
Over the course of the campaign, he or she will also become deeply aware of their individual concerns, %
and through them, he or she will learn about the concerns %
of particular segments of the voter population that he or she hopes to serve. %
Those individuals to whom he or she is accountable are vested not only with that responsibility, %
but they are also invested with the time, space, and resources %
to learn about their actions and the challenges and incentives they responded to with them. %
They would develop a second-hand, but nevertheless meaningful understanding %
of the politician's complex world, even while maintaining a different base of operation %
and not being subject themselves to the same forces and incentives. %
Instead they would observe them from their own positions, %
subject to an independent and very different set of social dynamics. %

The politician (and others) might try to influence the distribution of valence, %
or make other attempts to de-platform or weaken %
journalists who are less likely to support them. %
However, journalists would immediately recognize these ploys and would (perhaps, in most cases) %
support one another and act to warn voters, acting as a buffer against the effectiveness of these tactics. %
If these ploys are not effective enough, then the politician may have no recourse %
but to act as the system originally intended, and work in good faith to win the highest level of approval. %
The incentive for the politician to speak in reality-distorting ways, %
particularly during the immediate run-up to the election, when voters have already %
selected their journalistic proxies, would be drastically changed. %
It is difficult to say precisely what the effects would be on the complex dynamics %
of a political campaign; however, we can say that the overall effect would likely be to moderate %
and regulate the campaign process. %

{\em In the Journalistic Voting System, the politician's burden is eased.} 

\end{enumerate}
We conclude that in the Journalistic Voting System, %
the burden of each participant (voter, journalist, and politician) in the voting system is eased, %
relative to a traditional democratic system. %
In other words, each participant is happier on a relative basis, and has a better quality of life. %
This conclusion is based on a theoretical view. It is not known %
whether these claims are true empirically. %

\subsection{Potential Problems with Journalistic Voting System}\label{s.jd.prob}

The following is a categorized discussion of some potential problems with the %
Journalistic Voting System, which is intended to serve as a basis for discussion and criticism, %
not by any means as a final word. %
The views reflected here may someday appear optimistic in hindsight; time may tell. %

\subsubsection{Would it require nationalizing journalism and the media?}
No, that is not what is suggested. %
There should be no reason to ``nationalize'' companies or %
make significant modifications to existing companies and organizations. %

On the contrary, it could even be hoped that such a system %
might strengthen private journalism and reverse sad recent trends, %
such as the decline and evisceration of local journalism %
and the collapse of newspapers. %
The system for public journalism would exist alongside that for private journalism, %
which would continue to provide broader services, such as entertainment and sports news. %
Public journalism could also be regarded as a path into the private journalism profession, %
helping to support and improve journalism at all levels and creating opportunities %
for journalists to get a start, or to acquire greater knowledge and experience. %
For private companies the enterprise as a whole could be viewed as %
a valuable and essentially free resource for training and development, %
and a mine for extracting top talent. %

\subsubsection{Would any companies be shut down or regulated?}
Certainly there is no requirement whatsoever that any business be shut down. %
Perhaps some new regulations would be deemed necessary. %
However, the proposed role of the public journalist is very similar to roles %
that are already well known and well understood, such as: public servant, public office holder, %
member of the state bureaucracy. %
The impact would be likely to be {\em more} like creating a new department in the state bureaucracy, %
and {\em less} like re-ordering a sector of the already existing economy. %

\subsubsection{Is it antidemocratic?}
If we understand an antidemocratic system %
to be a system in which power and control is exercised without accountability or oversight, %
then the answer is no, because there is accountability between voters and journalists, %
who continually stay in close contact with one another. %
Indeed, journalists and voters are arguably {\em uniquely} appropriate for this connection of accountability, %
because a journalist, who is naturally candid and comfortably spends a significant part of his or her time in the public eye %
sharing in great detail his or her own views, knowledge, experience, sense of humor, etc., %
has as close and as meaningful of a relationship as it is possible for %
a large (million or more) population to have with an individual. %

\subsubsection{Is it secure?}
Security is an issue for any voting system. We can consider a small number of examples of possible security breaches. %
\begin{itemize}
\item 
A journalist is paid to vote a certain way, or votes a certain way for intangible benefits %
that are not faithfully represented to his or her followers and subscribers. %
\end{itemize}
To the extent that this is a security issue and not a political issue, %
we can see that {\em we have similar security problems in other election systems.} %
For example, a politician can be paid (bribed) to vote a certain way or an official can be paid %
(bribed) to take a certain action, or not to do so. %
As far as bribery or solitation of a bribe is concerned, %
such activities by a public journalist would constitute criminal activity, %
and if discovered a journalist would face conviction. %
\begin{itemize}
\item 
A voter is solicited to set valences a certain way (say, to support his friend who is a local journalist). %
\end{itemize}
Local politics is unfortunately plagued with friendly favors of this type. %
These are problems that one must face in a democracy, regardless of what the voting system is, %
and we already have methods to confront them. %

\subsubsection{Is it tested?}
The Journalistic Voting System {\em has not been tested.} %
Its performance at small scales, and its performance at larger scales, should all be tested and measured carefully %
to ensure that theoretically predicted performance matches empirical performance, %
before wagering the loss of social well-being against the performance of a new, untested voting system. %

\section{Conclusion}\label{s.conc}

We have reviewed the Valence Voting System and introduced the Journalistic Voting System. %
We noted that the Valence Voting System presents many compelling advantages, %
but at the same time, suffers from major drawbacks. %
In particular, the Valence Voting System resolves the problem of Gerrymandering, %
but it suffers from having unpredictable dynamics. %

Next, we saw that the Journalistic Voting System addresses the drawbacks %
that were observed with the Valence Voting System. %
We concluded, in addition, that the Journalistic Voting System %
presents several compelling advantages. %
First, it mitigates widely acknowledged threats to democracy from %
cybersecurity, social media, big data, artificial intelligence (AI), and gerrymandering. %
Second, we argued that it has the potential to enhance the %
happiness and effectiveness of participants who play roles in the system: %
politicians, journalists, and voters. %
In effect, it presents many of the advantages of the Valence Voting System, %
while avoiding its flaws. %
Moreover, it has the advantage of bearing a strong resemblance %
to the empirically-based world of present-day politicians and journalists, %
so it draws certain benefits from the historical, political, and sociological understanding %
of these long-standing dynamics. %

Several issues were noted with the proposed systems. %
The main problem noted with both systems is that neither has been empirically %
tested, and their true empirical properties at any scale are not currently known. %
Further research and testing would resolve this gap in our understanding. %
We are very fortunate to live in a world today where such organized, supervised testing would be %
possible, thanks to the cooperation, curiosity, communication, good faith, and understanding of the %
people who live in a modern, advanced democracy. %

In conclusion, we have argued that there are two basic reasons to consider the %
Journalistic Voting System: %
first, it might mitigate and be a deterrent against present threats to election security, %
and second, that it might be a better-functioning system for democracy's participants. %
While the latter point, in our view, deserves to be considered, it is the former point, not the latter, %
that presses the case and, in our view, calls the loudest for more thorough scientific investigation and analysis %
of the system. %

\bibliographystyle{abbrv}
\bibliography{jvs}

\end{document}